\documentclass[final,epsfig,amsfont]{article}
\usepackage{latexsym}
\usepackage[dvips]{graphicx}

\begin{document}

\title{The Multiparticle Quantum Arnol'd Cat: a test case for the decoherence approach to quantum chaos}

\date{}
\author{ Giorgio Mantica  \\ {\em International Center for
Non-linear and Complex Systems}\\ Universit\`a dell'Insubria, Via
Valleggio 11, Como, ITALY \\ and CNISM unit\`a di Como, \\ and  I.N.F.N. sezione di Milano.
}
\maketitle

\begin{abstract}
A multi-particle extension of the Arnol'd Cat Hamiltonian system is
defined and examined. We propose to compute its Alicki-Fannes quantum
dynamical entropy, to validate (or disprove) the validity of the
decoherence approach to quantum chaos. A first set of numerical
experiments is presented and discussed.

\end{abstract}

\section{Introduction}

In the phenomenological approach to decoherence, one defines a Markov map for the evolution of the reduced density matrix of the system. This map is expected to be an approximation of the true dynamics of a system interacting with an environment. Indeed, it can often be derived via some simplifying assumptions: typically in fact, like {\em e.g.} in the Caldeira--Legget model, the environment consists of a system with infinitely many degrees of freedom, whose exact dynamics is too complicate to be dealt with exactly.

This approach has proven to be very effective: one can list among its most important successes the explanation of the puzzling behavior of quantum Schr\"odinger cats \cite{haro1} and the emergence of classical properties in quantum mechanics \cite{joos}. For these reasons, a few years ago, the idea was proposed that decoherence might also explain the problem of quantum chaos, that is, it could restore dynamical chaos in the quantum evolution, that was missing in pure systems evolving via the Schr\"odinger equation \cite{tomas,andrey1,zurpaz}. Actually, this approach has a longer history: as early as 1984 Guarneri \cite{italo1} had shown that allowing a random perturbation in the dynamics of a kicked rotor washed away the phenomenon of quantum localization of classical chaos and Ott {\em et al.} \cite{ott} had worked out the parameter dependence of the ensuing diffusion coefficient $D$.
Dittrich and Graham \cite{tomas}, Kolovsky \cite{andrey1} and later Sundaram et al. \cite{arje1} have shown that the classical and quantum dynamics of a system, whose coupling to the environment leads to diffusion with coefficient $D$, are similar whenever $\sqrt {D/ \lambda}$ (here, $\lambda$ is the Lyapunov exponent) is larger than $\hbar$ times a dimensional constant. Indeed, in the presence of diffusion, the finest resolution discernable in phase space, $\delta p_{\min}$ scales as $\sqrt {D/ \lambda}$. Therefore, the conventional explanation goes by saying that when the latter quantity is larger than the discreteness of quantum phase space, of size $\hbar$, quantum effects do not have room to show up.
In the quoted references the similarity between the two dynamics has been exhibited by comparing classical phase--space distributions and Wigner functions, first visually and then by using a suitable distance function.

The system that has been most frequently analized in these investigations is the celebrated Arnol'd cat map \cite{avez,berry}, on which the failure of the correspondence principle (properly defined \cite{physd}) has been exhibited. Therefore, the following is a tale of two cats, Schr\"odinger versus Arnol'd: are the properties of decoherence exhibited by the former enough to assure the strong chaotic signatures of the second? The fate of the decoherence approach to quantum chaos lies in the answer to this question.

It must be remarked from the start that it is generally believed that the question above must be answered in the positive. In fact,
the theory relying upon effective equations for the density matrix mentioned above is certainly fruitful and physically motivated--as testified by an ever increasing list of works--and seems to support this point.
Yet, in our view, two points need to be investigated more deeply before the final word be said.

Firstly, the master equation approach cannot provide a complete answer to the problem of quantum--classical correspondence, no more than a phenomenological theory of fundamental interactions can be a complete field theory. Rather, one would like to have a fully dynamical, fully quantal calculation exhibiting the same phenomena of which the phenomenological approach is an approximation \cite{giubor}. This theory should also render evident the scaling relations in the physical parameters by which the limiting approximations are validated.

Secondly, we believe that this theory should address a very specific kind of questions. Indicators of some sort of irregularity abound in the field if quantum chaos \cite{karol,todd,giuli,giulii,lesh}, each possessing its relevance and limitations. Yet, we claim that since the essence of chaos is deterministic randomness (see \cite{joe} for a readable argumentation of this point), its foremost indicators must be information--theoretical. Moreover, if the classical limit is concerned, they must necessarily refer to the dynamical structure of phase space \cite{kosl,michael}.

In previous works \cite{physd,algocom2,algocom1,viva} we have used algorithmic complexity to gauge chaos in dynamical systems. In this paper, we rely on an equivalent quantity, the Alicki Fannes quantum dynamical entropy, hereby referred to as A-F \cite{alifa,alik} that offers the advantages of being computable to a certain extent, of arising from the consistent histories formalism \cite{histo} and of translating in the quantum domain the same construction procedure of the classical Kolmogorov Sinai entropy.

While the A-F entropy of systems with finite--dimensional Hilbert spaces is null, the scaling of the finite time Shannon entropies required for its construction is nonetheless informative.
We shall compute this quantity for a quantum Arnol'd cat that encounters multi--particle scattering. Indeed, that scattering by small particles can provide a fully dynamical model of decoherence has been already shown by Joos and Zeh \cite{joos}: we shall only render this model explicit and computable in a form that well ties with the classical model for the Arnol'd cat map, that can be seen as the Floquet evolution of a kicked particle on a torus.

As noted above, many works have already appeared linking decoherence and chaoticity, so that any list of references, including ours \cite{ian1,monpaz,alik0,marcos,zurek0,todd} is forced to be utterly partial. Nonetheless, we believe that the two theoretical points raised above have never been studied together so far. We therefore present in this paper the first results of our analysis, via information--theoretical dynamical entropies, of a fully dynamical model of a decohering system. Further study, both from the theoretical and the numerical point of view, will follow.

In the next section we review the classical Arnol'd cat and its quantization. Then, in Sect. \ref{sec-mpac} we introduce the multi--particle Arnol'd cat, in its classical and quantum versions. The problems involved in deriving a scattering matrix for quantum particles on a torus are solved in Sect. \ref{sec-scat}. Next, in Sect. \ref{sec-af} we review the essential ingredients of A-F entropy, particularly the  finite time Shannon Alicki Fannes entropy, that is then computed numerically in Sect. \ref{sec-resu} for the multi--particle Arnol'd cat. Its scaling with the system parameters is examined and it is discussed in the Conclusions.

\section{Single Particle Arnol'd Cat}
\label{sec-spac}

Consider a point particle of mass $M$ subject to move on a ring. Mathematically, this ring is a torus, of length $L$, labelled by the variable $Q \in [0,L)$. In the absence of any interaction, this particle rotates with constant velocity around the ring. To the contrary, it is also subject to the action of a periodic impulsive force, of period $T$, that has the effect of changing instantaneously its momentum: the relative Hamiltonian is therefore
\begin{equation}
 H_{cat}(Q,P,t) = \frac{P^2}{2M} -  \kappa \frac{Q^2}{2} \sum_{j=-\infty}^\infty \delta(t/T-j),
 \label{eq-ham1}
\end{equation}
$P$ being the conjugate momentum to $Q$ and $\kappa$ a coupling constant. This is the Hamiltonian for the Arnol'd cat mapping.
In fact, it entails a classical period evolution operator that acts as follows on the dynamical variables observed at the time immediately following the action of the impulsive force:
\begin{equation}
  \left\{
\begin{array}{l}
Q \rightarrow  Q +  \frac{T}{M} P    \\
P  \rightarrow P + \kappa T Q .
\end{array} \right.
\label{eq-ham2}
\end{equation}
It is now convenient to rescale momentum $P$ by multiplying it by $T/M$. The new momentum, $\tilde{P}:=  \frac{T}{M} P$ has the dimensions of a length. In the new variables, the action of the period evolution operator becomes:
\begin{equation}
  \left\{
\begin{array}{l}
Q \rightarrow  Q +  \tilde{P}  \\
\tilde{P}  \rightarrow \frac{\kappa T^2}{M} Q + (1+\frac{\kappa T^2}{M}) \tilde{P} .
\end{array} \right.
\label{eq-ham3}
\end{equation}

In addition, we stipulate that the rescaled momentum variable $\tilde{P}$ is also periodic, of the same period $L$ as of $Q$, so that classical dynamics evolve in a two--dimensional torus.
It is easy to see that the Hamiltonian map (\ref{eq-ham3}) can be made consistent with this geometry by choosing $L=1$ and $\frac{\kappa T^2}{M}$ to be an integer. Indeed, letting
\begin{equation}
\frac{\kappa T^2}{M}  = 1
\label{eq-ham4}
\end{equation}
exactly yields the renown classical Arnol'd cat on the two--dimensional torus. We shall always assume this condition in the following.

The quantization of the Hamiltonian (\ref{eq-ham1}) have been performed in \cite{berry} with semiclassical means and canonically in \cite{physd} to which we refer for additional comments. Rigorous mathematical work has clarified the generality of the kinematical procedure \cite{isola}, while physics stands as in the original work of Schwinger \cite{schw} and is indeed quite elementary. In fact, periodicity in the $Q$ variable implies that that wave--functions can be written in the form
 \begin{equation}
  \psi (Q) = \sum_k c_k \phi_k(Q)
  \label{eq-ham5}
\end{equation}
where $\phi_k(Q) := e^{-i 2 \pi k Q/L}$ are the momentum eigenfunctions and $c_k$ are the expansion coefficients, with integer $k$.
We need next to impose periodicity in $P$ with period $M L/T$.  Recall that the momentum operator is
$\hat{P} = i \hbar \partial_Q$, so that $\hat{P} \phi_k(Q) = \frac{2 \pi k \hbar}{L} \phi_k(Q)$ and the eigenmomentum is $h k / L$. Therefore, going to the momentum representation we must have that
\begin{equation}
c_k = c_{k+N},
\label{eq-ham6v}
\end{equation}
for any $k$, where $N$ is an integer number satisfying
\begin{equation}
\frac{M L^2}{T}  = N h,
\label{eq-ham6}
\end{equation}
and obviously $h$ is Planck's constant. We may certainly set $L=1$ here and in the following, with no loss of generality, since one parameter can be freely chosen. Eq. (\ref{eq-ham6}) has a physical origin: in order to realize quantum dynamics on a two dimensional torus of unit periodicity in the $Q$ direction, the periodicity in $P$ must be an integer multiple of the Planck constant.
As a consequence of eq. (\ref{eq-ham6v}), letting $c_k$ = $c_{k+N}$ in the expansion (\ref{eq-ham5}) factors out a periodic train of delta functions at the spatial locations $Q_j = \frac{j}{N}+s$, with $j=0,\ldots,N-1$, and $s$ a fixed shift. Wave--functions can therefore be represented as vectors in a finite--dimensional Hilbert space of dimension $N$, that can be formally expressed in a convenient normalization as
 \begin{equation}
  \psi (Q_j) = \frac{1}{\sqrt{N}} \sum_{k=0}^{N-1} c_k \phi_k(Q_j) =
  \frac{1}{\sqrt{N}} \sum_{k=0}^{N-1} c_k e^{-i 2 \pi k j/N}.
  \label{eq-ham5b}
\end{equation}
 The dual momentum representation is easily obtained by a discrete Fourier transform of the former: letting $P_k = k h$, this is
\begin{equation}
  c_k = \hat{\psi}(P_k) = \frac{1}{\sqrt{N}} \sum_{j=0}^{N-1} \psi(Q_j)e^{i 2 \pi k j /N}.
  \label{eq-ham5c}
\end{equation}

Observe finally that the dimension of the Hilbert space is directly proportional to the mass $M$ of the particle. This is in line with our approach of performing the classical limit in its more transparent physical form, by keeping $\hbar$ to its real physical value and by considering a particle of larger and larger mass $M$.

By standard quantization procedures one therefore compute the matrix representations of the evolution operator. This has been effected in \cite{physd}. We reproduce here the formulae holding in the position representation, where the state vector has components $\psi(Q_j)$, $j=0,\ldots,N$. The action of the free evolution $U^{free}$ induced by the free rotation $\frac{P^2}{2M}$ has matrix elements
\begin{equation}
U^{free}_{kl} = \frac{1}{\sqrt{N}} e^{- (\pi i l^2 /N)} e^{2 \pi i kl /N}.
  \label{eq-rota1}
 \end{equation}
The quantum Arnol'd cat evolution operator is the product $U^{cat} = K U^{free}$, where $U^{free}$ has been defined just above and $K$ is the operator with matrix elements
\begin{equation}
K_{kl} = \frac{1}{\sqrt{N}} e^{ i \pi l^2 / N} \delta_{k,l},
  \label{eq-gatto1}
 \end{equation}
where $\delta_{k,l}$ is the Kronecker delta,
that corresponds to the impulsive part of the Hamiltonian.

\section{Multi--particle Arnol'd Cat}
\label{sec-mpac}

We now consider a more complex system consisting of a single large particle of mass $M$ and of a number $I$ of smaller particles of mass $m$, that are also bound to move on the same ring. Let $q_i$ and $p_i$, $i=1,\ldots,I$ the coordinates and momenta of these particles, respectively. We require that also the phase space of the small particles is a two dimensional torus, of $L=1$ periodicity in the variables $q_i$ and of periodicity $\frac{mL}{T}$ in the momenta $p_i$. We must therefore have
\begin{equation}
   \frac{m}{T} = n h,
  \label{eq-ham5a}
\end{equation}
where $n$ is an integer. The wave--functions of the small particles also take on the form of eqs. (\ref{eq-ham5b},\ref{eq-ham5c}), with $n$ in place of $N$.
The many--particles wave--functions can be written on the basis of momentum eigen--functions as
\begin{equation}
  \Psi (Q,q_1,\ldots,q_I) = N^{-1/2} n^{-I/2} \sum_{k=0}^{N-1}
   \sum_{k_1=0}^{n-1} \cdots
   \sum_{k_I=0}^{n-1}
   c_{k,k_1,\ldots,k_I} e^{-2 \pi i(k Q + \sum_i k_i q_i)}.
  \label{eq-ham5m}
\end{equation}

Notice that the coordinates $q_i$ are restricted to a lattice of spacing $1/n$, containing $n$ points.  Similarly, the momenta $p_i$ live on the lattice $k h$, with $k=0,\ldots,n-1$.

In this paper, we choose $N$ to be an integer multiple of $n$ (that is, $N = p n$, that also means that the large mass $M$ is a multiple of the small mass $m$, {\em i.e.} $M=pm$). We also impose that the position--momentum lattice of a small particle is a subset of that of the large one: to achieve this, we write the position lattice of the $i$-th particle in the form
 \begin{equation}
  q_i = j \frac{1}{n} + s_i \frac{1}{N}, \;\; j =0,\ldots,n-1.
  \label{eq-lat1}
\end{equation}
where $s_i$ an integer measuring the shift of the position lattice of the $i$-th particle with respect to that of the large particle. The allowed values of the constants $s_i$ range from zero to $N/n -1$.
Therefore, in the position representation the state of the system is represented by the values of $\Psi$ at
$(Q,q_1,\ldots,q_I) = (\frac{j}{N},\frac{j_1}{n}+s_1/N,\ldots,\frac{j_I}{n}+s_I/N)$, that we label as $\Psi_{j_0,j_1,\ldots,j_I}$ using the index zero for the large particle.
Mapping to and from the two representations is easily effected via the multi-dimensional discrete Fourier transformation.

Having taken care of kinematics, let's focus on dynamics: the   Hamiltonian of the multi--particle Arnol'd cat is
\begin{equation}
 H = H_{cat}(Q,P,t) + \sum_{i=1}^I \frac{1}{2m} p_i^2 + V \sum_{i=1}^I
  \Phi(q_i-Q),
 \label{eq-ham10}
\end{equation}
where $V$ is a coupling constant and the function $\Phi$ will be described momentarily. Small particles also rotate freely on the torus, except for an interaction potential $\Phi$ with the large particle. The form of the interaction translates elastic scattering between the large particle and each of the smaller ones. In turn, these latter do not interact among themselves. It is easy, although not necessary at this stage, to introduce also an interaction among the small particles.
The form of the Hamiltonian (\ref{eq-ham10}) follows the decoherence program of Joos and Zeh \cite{joos}: the large particle encounters frequent collisions with the small ones, that should ultimately result into decoherence and classicality.

\section{The scattering matrix elements}
\label{sec-scat}

We must now make precise the interaction potential $\Phi$ appearing in the Hamiltonian (\ref{eq-ham10}). Our goal would be to have a hard core potential equal to a Dirac delta function, representing a perfectly elastic scattering \cite{teta1}. Yet, we have to cope with the fact that kinematics takes place in the tensor product of quantized two--dimensional tori. On the one hand, this makes the problem easily solvable, as we now show. On the other hand, it yields significant new features in the scattering process which will be described elsewhere.

Since we are considering only the interaction between the large particle and each small one independently, we can write the interaction potential $\Phi$ in the form
\begin{equation}
  \Phi = \sum_{i=1}^I \Phi^{(i)} \otimes I^{(i)},
  \label{eq-scat1}
 \end{equation}
where
$\Phi^{(i)}$ is the interaction matrix in the $(0,i)$ subspace (for convenience of notation, we shall also let the index $0$ label the position--momentum of the large particle: $q_0:=Q$, $p_0=P$) and
$I^{(i)}$ is the identity in the orthogonal complement of the $(0,i)$ subspace.
The interaction between the large (zeroth) particle and the $i$-th small one is effective only when they sit on the same lattice point: it is therefore convenient to
specify the scattering potential directly in the position representation, where $\Psi$ is defined by its values at the lattice positions
$(Q,q_1,\ldots,q_I) = (\frac{j}{N},\frac{j_1}{n}+s_1/N,\ldots,\frac{j_I}{n}+s_I/N)$.
In this basis we have
\begin{equation}
  \Phi^{(i)}_{l_0,k_i;l'_0,k'_i} = \delta_{l_0,l'_0} \delta_{k_i,k'_i} \delta_{l_0,p k_i + s_i},
  \label{eq-scat2}
\end{equation}
where $p=\frac{N}{n}$, where $l_0$ and $l'_0$ range from $0$ to $N-1$, while $k_i$ and $k'_i$ range from $0$ to $n-1$. As anticipated, the last Kronecker delta requires that the particle $0$ and $i$ sit at the same lattice point. According to eqs. (\ref{eq-scat1}) and (\ref{eq-scat2}), the full matrix elements of $\Phi$ are therefore
\begin{equation}
  \Phi_{l_0,k_1,\ldots,k_I;l'_0,k'_1,\ldots,k'_I} =
   \delta_{l_0,l'_0} \prod_{i=1}^I \delta_{k_i,k'_i}
   \sum_{i=1}^I \delta_{l_0,p k_i + s_i}.
  \label{eq-scat3}
\end{equation}
It is therefore apparent that $\Phi$ is diagonal in the coordinate representation.

Finally, the form of the Hamiltonian (\ref{eq-ham10})  suggests a numerical technique for the quantum evolution. Write symbolically \begin{equation}
  H = H^{free} + \Phi^{scat} + K \sum_{j=-\infty}^\infty \delta(t/T-j),
  \label{eq-scat4}
\end{equation}
where $H^{free}$ is the free motion Hamiltonian, $\Phi^{scat}$ is the scattering contribution, and $K$ is the impulsive force (acting only on the $Q$ coordinate). Then, the full period evolution operator $U$ can again be written as the product of
$U_0 := e^{-i \hbar T (H^{free} + \Phi^{scat})}$ and of $U_{kick} := e^{-i \hbar T K}$.
In turn, the former can be evaluated as a Trotter product form
 \begin{equation}
   U_0 = \prod_{r=0}^{R-1} e^{-i \hbar \frac{T}{R} H^{free}}
    e^{-i \hbar \frac{T}{R} \Phi^{scat}},
  \label{eq-scat4}
\end{equation}
whose accuracy increases by increasing the integer number of partitions $R$ of the interval $(0,T)$. Numerical convenience demands finally that each exponential be evaluated in the basis where the corresponding operator is diagonal. This is swiftly accomplished by tangling the products above with fast Fourier multidimensional transformations. The full code has been programmed in stone--age Fortran 77 language, of which the author is a proud cultivator.

\section{S-A-F quantum dynamical entropy}
\label{sec-af}

The essential ingredient of defining a dynamical entropy is coarse--graining, that leads to symbolic dynamics.
Suppose that the Hilbert space of the system is partitioned into cells, corresponding to projection operators $P_k$, so that their sum  is the identity operator: $\sum_k P_k = I$. Given any initial state
$\psi$, quantum evolution yields the vector $\psi(j)$ at any future (or past) time $j \tau$, with $j \in \mathbf Z$ and $\tau$ an observation delay. Clearly, the probability that the quantum system is found in macro--state $k$ at time
$j$ is given by the square modulus of $P_k \psi (j)$.  The
``quantum history'' of a vector $\psi$ is
the result of repeated projections on macro--states, followed by unitary evolution. If the choice of the macrostate at time $j$ is indicated by $\sigma_j$, and if the string of choices at successive times is entered in the vector $\mathbf{\sigma}$ of length $J$,
$\mathbf{\sigma}=(\sigma_0,\sigma_1,\ldots,\sigma_{M-1})$ (this is called a ``word'' in symbolic dynamics) the quantum history of the vector $\psi$ is
\begin{equation}
  \psi_\sigma =  (U P_{\sigma^{}_{J-1}}) \cdots (U P_{\sigma^{}_0}) \psi.
  \label{proj2}
 \end{equation}
For convenience of notation we shall put
 \[
U^{\sigma_j} := U  P_{\sigma_j}, \;\;\; j = 0,\ldots,J-1.
 \]
The ``amplitude'' $(\psi_\sigma, \psi_\sigma)$, when averaged over initial conditions $\psi$, as we shall do momentarily, is the analogue of
the measure of the classical phase space cylinder associated with
the symbolic dynamics $\sigma$. The formal analogy is completed by noting the equivalence of $U$ with the inverse of the classical map. In both
classical and quantum dynamics these probabilities add up to one:
$\sum_\sigma (\psi_\sigma, \psi_\sigma) = 1.$ In quantum mechanics,
though, interference reigns and the products $(\psi_\sigma,
\psi_{\sigma'})$ are non-null also when $\sigma \neq \sigma'$.

Kolmogorov--Sinai entropy is constructed starting from the measures of the cylinders $\sigma$. In the
A-F. quantum formulation \cite{alik}, entropy is derived by the
spectrum of the decoherence matrix $D$ with entries
$D_{\sigma,\sigma'}$, defined by
\begin{equation}
 D_{\sigma,\sigma'} := \frac{1}{\cal N}
 Tr ( U^{\sigma^{}_{J-1} \dagger} \cdots U^{\sigma^{}_0 \dagger}
U^{\sigma'_0} \cdots U^{\sigma'_{J-1}} ),
  \label{proj3a}
\end{equation}
where ${\cal N}$ is the dimension of the Hilbert space, the dagger indicates the adjoint and clearly $U^\dagger =
U^{-1}$, $P_k^\dagger = P_k$. Observe that $D$ is a $2^J \times 2^J$
square matrix, Hermitean, of unit-trace and non-negative. In the classical case, this matrix is diagonal. In the quantum case, one defines the Shannon - Alicki - Fannes (S-A-F) entropy $S(J)$ of the system histories of length $J$ with projections
$\{P_k\}$  as
\begin{equation}
   S(J) = Tr ( - D \log D).
  \label{proj3}
 \end{equation}

Technically, the A-F entropy associated with the partition  $\{P_k\}$ is the limit $S(J)-S(J-1)$ as $J$ tends to infinity, as in the case of KS entropy. For systems with finite--dimensional Hilbert spaces it is null. Nonetheless we ascribe particular importance to the S-A-F entropies $S(J)$ even before the limit is taken and the required supremum over partitions is effected. In fact, these finite-partition, finite-time entropies, {\em and their scaling behavior} with respect to the system parameters are in our view the most significant physical quantities.

The numerical problem of computing the quantity $S(J)$ for systems with a finite dimensional Hilbert space has been discussed in \cite{etna}. One consider a different matrix, also introduced by A-F, with the same spectral properties as $D$, but with a size that is independent of the word length $J$. Relying on the properties of this latter matrix, derived in \cite{etna}, a general purpose parallel code has been designed and will be employed in this paper.

\section{The A-F entropy of the multi-particle Arnol'd cat}
\label{sec-resu}

Let us now consider a partition of classical phase--space in four equal cells of rectangular shape, defined by letting the position of the large particle, $Q$, belong to the sets $[k/4,(k+1)/4)$. In these cells the momentum $P$ of the large particle and the coordinates (positions and momenta) of the small ones take on all the allowed values. This partition can be easily generalized, but there is no need to do that in the present context. Given this partition, when only the large particle is present, classical theory provides us with KS entropy of the Arnol'd cat map.

\begin{figure}[h]
\center
\includegraphics[width=8.5cm,height=12cm,angle=270]{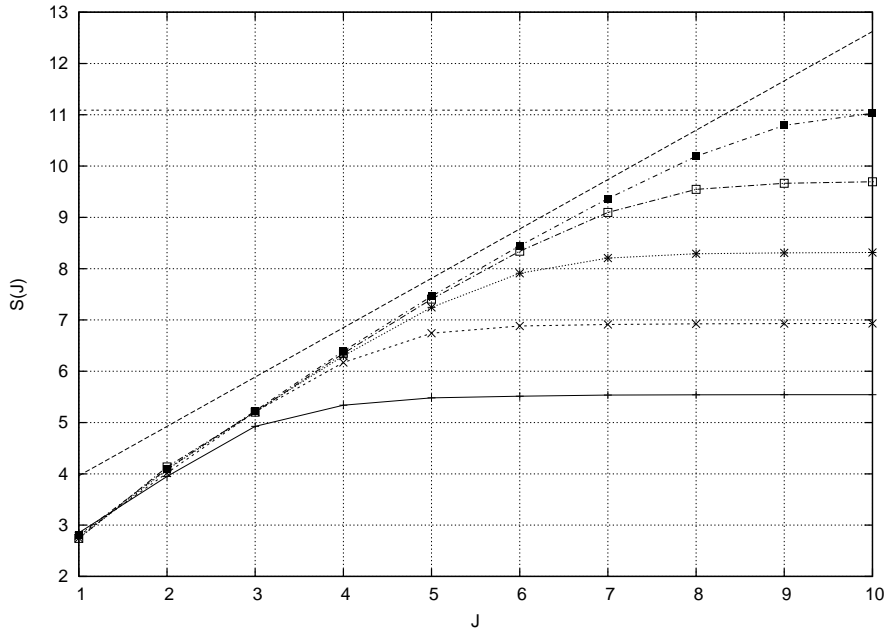}
\caption{S-A-F entropies $S(J)$ for the one--particle Arnol'd cat mapping, with $N=2^4$ (continuous line, pluses), $N=2^5$ (dashed line, crosses), $N=2^6$ (dotted line, asterisks), $N=2^7$ (dotted-dashed line, open squares) and $N=2^8$ (dotted-dashed line, full squares). Also drawn are the horizontal line at eight $S = \log 2^{16}$ (theoretical limit for $S(J)$ in the last case) and a line with slope equal to the K-S entropy of the classical Arnol'd cat.}
\label{fig-incdim}
\end{figure}

Corresponding to each partition cell, we can define a projection operator in the Hilbert space of the system, whose form is trivially simple when the wave--function $\Psi$ is written in the position representation. We can therefore compute the S-A-F entropies $S(J)$. Let us start from the single--particle case, that has already been described in \cite{etna} and, just for the linear entropy, a quantity that can be computed more easily than Shannon's,  in \cite{alimoni}. In
Fig. \ref{fig-incdim} we plot these functions for increasing values of $N$. We observe that as $N$ increases, the linear behavior (and the numerical values) of the classical cat is approached, for a region in $J$ of increasing size. This region terminates as soon as the linear increase of $S(J)$ is hampered by the finiteness of Hilbert space, via the bound $S(J) \leq 2 \log({\cal N})$. Here, ${\cal N} = N$. This bounds translates on the one side the minimal size of phase--space cells implied by quantization and on the other side the finite amount of algorithmic information content of the quantum motion.

Therefore, fig. \ref{fig-incdim} is just another mathematical confirmation of the thesis of ref. \cite{physd} (see also \cite{viva}). It is there pretended that the correspondence principle is physically irrelevant for this system, on the basis of the simple observation that to achieve a {\em linear} increase of the time-lag of chaotic behavior (the region with information production, {\em i.e.} increasing $S(J)$), an {\em exponential} increase of ${\cal N}$ (and therefore of the mass $M$, which is proportional to the former, see the formulae in Sect. \ref{sec-spac}) is required.

To the contrary, for the multi--particle Arnol'd cat map, the bound is not so stringent: in fact, the full Hilbert space ${\cal H}$ of the system is the tensor product of the single particle Hilbert spaces:
\begin{equation}
 {\cal H} = H_{big} \otimes H_{small}^I,
  \label{eq-gatto1}
 \end{equation}
so that the dimension of  ${\cal H}$ is ${\cal N} = N n^I$, and the logarithmic upper bound to complexity can be increased linearly by a realistic, linear increase of the number $I$ of particles: to paraphrase Saunders Mac Lane, {\em Gentlemen: there is lots of room in THIS Hilbert space}.

Certainly, this is only necessary and not sufficient to insure that the multi-particle quantum Arnol'd cat will physically behave in the limit as its classical relative. The next (but not final) requirement is that the unorganized information present in the reservoir of small particles be organized by the dynamics in order to produce  information at the classical rate given by the Kolmogorov-Sinai entropy.

\begin{figure}[h]
\center
\includegraphics[width=8.5cm,height=12cm,angle=270]{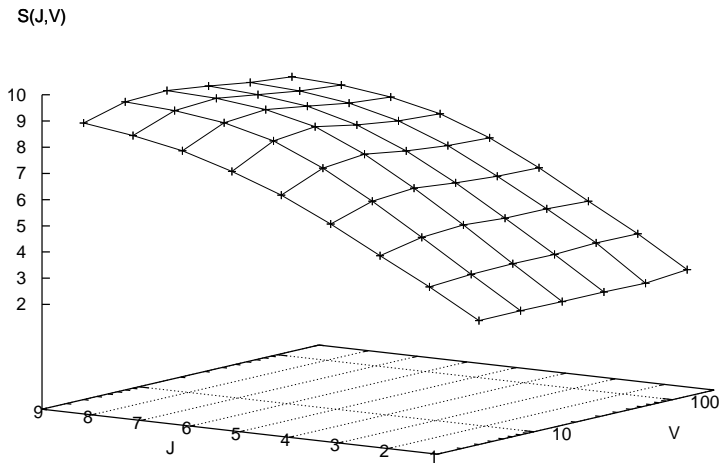}
\caption{S-A-F entropies $S(J,V)$ for the multi--particle Arnol'd cat mapping versus $J$ and $V$. Parameters are $M=2^4h$,  $m=2h$, $I=3$.}
\label{fig-sjv}
\end{figure}
 
Let us therefore present the result of numerical experiments. We first show the effect of coupling to the small particles. 
Figure \ref{fig-sjv} displays the entropies $S(J)$ versus $J$ and $V$, the scattering coupling constant, for the case of a large particle of mass $M=2^4h$ interacting with 3 small particles of mass $2h$. Recall that the single particle Hilbert space dimension is equal to the mass of the particle divided by $h$.

The case $V=0$ corresponds to the lowest curve in Fig. \ref{fig-incdim}, that at $J=5$ has almost completely attained its limit value $8 \log 2 \sim 5.545$. We observe that, as $V$ grows, a significant range appears where $S(J,V)$ is approximately stationary in $V$, for fixed $J$. Notice that the scale in $V$ is logarithmic. The same behavior is to be found in other cases. Let us therefore choose a value of $V$ within this range and look at the dependence of $S(J,V)$ with $J$.

\begin{figure}[h]
\center
\includegraphics[width=8.5cm,height=12cm,angle=270]{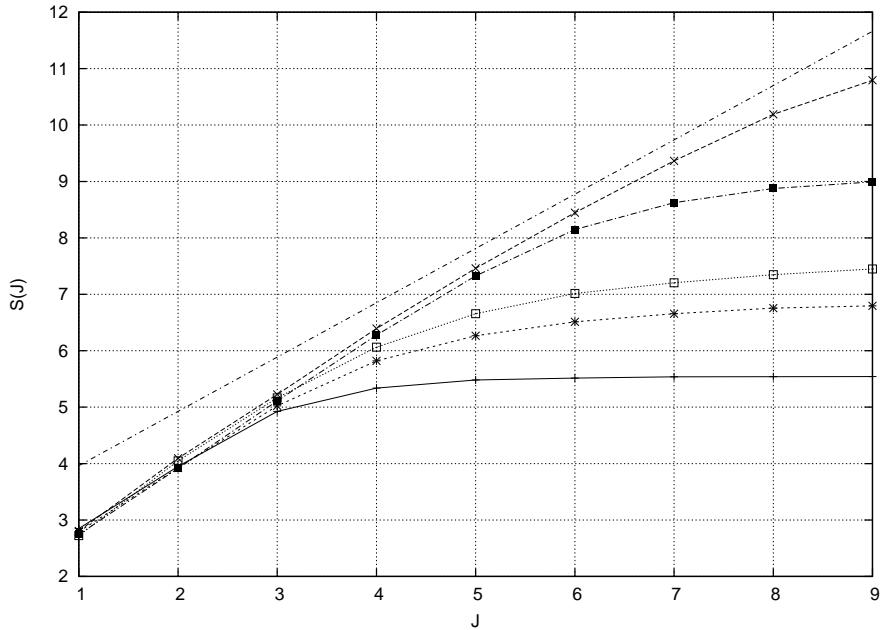}
\caption{S-A-F entropies $S(J)$ for the multi--particle Arnol'd cat mapping versus $J$. Parameters are: $V=8$, $M=2^4h$, $m=2h$ and $I=1$ (dashed line, asterisks), $I=2$ (dotted line, open squares), $I=3$ (dot-dashed line, full squares). Also reported for comparison are two sets of data from Fig. \ref{fig-incdim}, that is, $V=0$ and $M=2^4h$  (continuous line, pluses) and $M=2^8h$ (dashed line, crosses) }
\label{fig-dim4}
\end{figure}

If we do this, we observe that the function $S(J)$ increases far beyond what observed in Fig. \ref{fig-incdim}. To appreciate quantitatively this fact, we performed a sequence of numerical experiments by varying the number of small particles. In Fig. \ref{fig-dim4} we have kept $V=8$, $M=2^4h$, $m=2h$ and we let the number of small particles $I$ vary between one and four. For comparison, we have also plotted again the values of $S(J)$ without coupling for $M=2^4h$ and $M=2^8h$. The latter curve is a good approximation of the classical entropies for $J$ less than, or equal to six. We notice that in going from $I=1$ to $I=3$ the range of coincidence of $S(J)$ and $I$ with the classical curve increases steadily, even if $M$ is constant. 

It therefore seems that increasing the number of small particles up to an optimal value, at fixed masses $M$ and $m$, leads to an increase of the time--span of quantum classical correspondence, as far as information production is concerned. Moreover, this increase seems to be a power-law function (certainly not a logarithmic one) of the number of small particles.

In assessing the significance of these numerical experiments, one must notice that the variables involved have an exponential role. The number of particles, $I$, appears in the exponent of the Hilbert space dimension ${\cal N}$, and numerical computation of the Shannon entropies $S(J)$ has a computational cost that scales as a power of ${\cal N}$, depending on implementation \cite{etna}. The  runs for this paper have required thousands of cpu hours on large clusters of processors. Far from being a limitation of the technique, this computational complexity reveals the physical depth of the Alicki Fannes entropies, as discussed in \cite{etna}. It also explains why these quantum entropies have not (yet) received the attention they deserve, because they are so difficult to compute.

\section{Conclusions}

In this paper we have introduced a multi--particle version of the celebrated Arnol'd cat mapping. It was devised with the aim to study dynamically the validity of the decoherence approach to quantum chaos. We have reported preliminary data showing the path to be taken to decide the case. In fact, the following course of investigation lies now clear in front of us: one should first compute the classical Shannon entropies $S(J)$ for a generating partition of phase space and compare them with the A-F Shannon entropies for the same partition. Given a fixed precision $\varepsilon$ and a fixed mass of the small particles $m$, find pairs of parameters $M$ (mass of the large particle) and $I$ (number of small particles) so that the classical and quantum $S(J)$ coincide within precision $\varepsilon$ for all $J$ smaller than a maximal value $J_{max}$. At this point one should verify whether there exists sequences of $(M,I)$ that grow at most polynomially in $J_{max}$.

If this were to be the case, one could conclude that indeed the decoherence approach can justify on physical grounds the correspondence principle for chaotic systems, at least as far as the total information production is concerned. 

It is altogether clear that this endeavor is highly challenging, given the computational complexity described at the end of the previous section. Yet, our data (that although preliminary have required extensive computation and coding development \cite{etna}) seem to indicate that the effort might be successful.

In sequence, one should then consider the specific (as opposed to global) information production: one should investigate, under the same conditions just specified, whether individual histories become consistent, thereby reproducing the classical measures of dynamical cylinders. This remarkable fact has been shown so far only for under Markovian equations \cite{halli} and it seems to appear also in QSD models \cite{ian2}, but to our knowledge it has never been derived under ``controlled'' dynamical conditions such as those advocated in our approach.

All these investigations will be the matter of our future efforts.
It is finally worth to remark a related endeavor that points in the same direction: the similarity of a quantum system with a classical system on a lattice \cite{physd} has been confirmed by studying algorithmic information content and classical Kolmogorov Sinai Shannon entropies \cite{algocom2} \cite{bena}. If in addition a random noise is introduced, a regime in parameter space has been shown to exist so that the discrete system with randomness outputs information at the same rate as the classical continuous system \cite{disc1}.

\section{Acknowledgements}
The calculations of this paper have been produced on the CINECA parallel processor clusters BCX, CLX and SP6 thanks to the grant "Open quantum systems: quantum entropy and decoherence" for Progetti di Supercalcolo in Fisica della Materia, Project Key : giorgiomantica376572968172.

\end{document}